\documentclass[aps,prl,floatfix,twocolumn,reprint,groupedaddress,showpacs]{revtex4-1}
\usepackage{amsfonts}
\usepackage{mathrsfs}
\usepackage{amsmath}
\usepackage{color}
\usepackage{graphicx}
\usepackage{bm}
\usepackage{amssymb}
\usepackage{xspace}
\usepackage{epstopdf}
\usepackage{dcolumn}
\usepackage{longtable}
\usepackage{multirow}
\usepackage[colorlinks=true, letterpaper=true, pdfstartview=FitV, linkcolor=blue, citecolor=blue, urlcolor=blue]{hyperref}

\makeatletter

\newcommand{\Rmnum}[1]{\expandafter\@slowromancap\romannumeral #1@}
\makeatother

\begin{document}
\title{Unusual Magneto-Response in Type-II Weyl Semimetals}

\author{Zhi-Ming Yu}
\affiliation{Beijing Key Laboratory of Nanophotonics and Ultrafine Optoelectronic Systems, School of Physics, Beijing Institute of Technology, Beijing 100081, China}

\author{Yugui Yao}
\email{ygyao@bit.edu.cn}
\affiliation{Beijing Key Laboratory of Nanophotonics and Ultrafine Optoelectronic Systems, School of Physics, Beijing Institute of Technology, Beijing 100081, China}

\author{Shengyuan A. Yang}
\email{shengyuan\underline{\ }yang@sutd.edu.sg}
\affiliation{Research Laboratory for Quantum Materials, Singapore University of Technology and Design, Singapore 487372, Singapore}

\begin{abstract}
We show several distinct signatures in the magneto-response of type-II Weyl semimetals. The energy tilt tends to squeeze the Landau levels (LLs), and for a type-II Weyl node, there always exists a critical angle between the $B$-field and the tilt, at which the LL spectrum collapses, regardless of the field strength. Before the collapse, signatures also appear in the magneto-optical spectrum, including the invariable presence of intraband peaks, the absence of absorption tails,
and the special anisotropic field dependence.
\end{abstract}
\maketitle

The exploration of solids with nontrivial band  topologies has become a focus in current research~\cite{Kane,Qi}. Besides novel physical effects and application perspectives, the interest also comes from the possibility of simulating intriguing elementary particles phenomena in condensed matter systems. Notably, the Weyl fermion, which was originally proposed as a massless solution of the Dirac equation but remained elusive in high-energy experiments, could find its realization as low-energy quasiparticles~\cite{Wan,Murakami,Balents,Ran,Xu,Bernevig,Lu,Gong,Sau,Das,Yong1,Yang,Liu} in the so-called Weyl semimetals (WSMs). In a WSM, the conduction and valence bands touch with linear dispersion at isolated Fermi points known as Weyl nodes.
Each Weyl node is like a monopole in reciprocal space, carrying a topological charge of $\pm 1$ corresponding to its chirality. Weyl nodes of opposite chiralities appear or annihilate in pairs~\cite{no-go}, and at the system boundary their projections are connected by surface Fermi arcs~\cite{Wan}. The recent progress in identifying several WSM materials~\cite{ex1,ex2,ex3,ex4,ex5,ex6,ex7,ex8,ex9} have driven a flurry of exciting researches trying to probe the various fascinating phenomena connected to Weyl fermions~\cite{AV,Pesin,Fiete,Xie,YPZ,NN,Volovik,Aji,Son1,Son2,Zyu,Gru,Gos,Hos,Franz,Burkov-14,CX,Qi-optics,Voz,Zhou}.

The energy dispersion at a Weyl node could generally be tilted along  a certain direction in $k$-space. When the tilt is large enough, the Weyl cone could even be tipped over such that the Fermi surface transforms from a point to a line or a surface. Such Weyl nodes are referred to as type-II to be distinguished from the conventional ones, and have recently been proposed in a few materials~\cite{Yong2,Solu,Ruan,ZWang,XuSY,Kami,Autes}.
The essential topology (like chirality) of the Weyl node is unchanged by the tilt, however, since the geometry of Fermi surface plays a key role in many material properties, the type-II WSMs are expected to exhibit signatures distinct from the conventional WSMs and also other materials, e.g., as manifested in the predicted anisotropic chiral anomaly and anomalous Hall effects~\cite{Solu,Zyuz}.

Under an external magnetic field, electrons' motion is typically quantized into discrete Landau levels (LLs). In a three-dimensional (3D) solid, these LLs become dispersive in the direction along the field, such is the case also for conventional WSMs.
Here we show that
the additional energy tilt tends to squeeze the Landau level spacing, and remarkably, for a type-II node, the squeezing can be so dramatic that there always exists a critical angle between the $B$-field and the tilt direction, at which the LL spectrum collapses, regardless of the field strength. We provide a semiclassical picture for understanding such effects, showing that the collapse corresponds to a transformation of cyclotron orbits beyond the effective Weyl model. Before collapse, the transitions between LLs give rise to absorption peaks in the optical conductivity. For type-II nodes we find that these peaks exhibit unique features distinct from conventional WSMs and other materials, particularly for the processes involving the anomalous zeroth LL. These findings provide experimental signatures for type-II WSMs, and we also discuss possible ways for experimentally quantifying the tilt.

The essential physics that we describe in this work can be captured by the following simple $2\times 2$ Weyl Hamiltonian,
\begin{equation}\label{H0}
\mathcal{H}= v_0\bm k\cdot\bm \sigma + \bm w\cdot \bm k\; \mathbb{I}_{2\times 2},
\end{equation}
where $\bm \sigma$ is the vector of Pauli matrices,  $\mathbb{I}_{2\times 2}$ is the identity matrix, $v_0$ is the Fermi velocity (its sign gives the chirality of the node, and for definiteness we take $v_0$ to be positive in the following calculation), and the second term with vector $\bm w$ denotes the tilt of spectrum. A finite $\bm w$ tilts the dispersion along $\hat{\bm w}$, where $\hat{\bm w}=\bm w/w$ is the unit vector along the tilt direction. For $(w/v_0)<1$, the Weyl node is  conventional with $k=0$ being the only zero-energy mode. However, when $(w/v_0)>1$, the linear dispersion cone along $\hat{\bm w}$ will be tipped over, and the node becomes type-II. In both cases, the Weyl node is topologically robust in that all the three Pauli matrices are used up hence any small perturbations can only shift the location of the node but cannot remove it.

{\color{blue}{\em LL Squeezing and Collapse.}}---Under an external magnetic field, we make the usual Peierls substitution $\bm k\rightarrow \bm k+e\bm A$ (we set $\hbar=1$ here) in Hamiltonian (\ref{H0}) with the vector potential $\bm A$. We neglect possible Zeeman splitting since it is typically much smaller than the orbital effect at accessible field strength.
Without loss of generality, we could choose our coordinates such that the $z$-axis is along the $B$-field and $\bm w=(w_\bot,0,w_\|)$ lies in the $x$-$z$ plane, where $w_\|=\bm w\cdot \bm B/B$ ($w_\bot=|\bm w\times \bm B/B|$) is the projection of the tilt along (perpendicular to) the $B$-field.
Using the gauge $\bm A=(-By,0,0)$, one observes that the tilt gives rise to a term $-ew_\bot By$, which is equivalent to the effect of an electric field $E_\text{eff}=w_\bot B$ along the negative $y$-direction.
Since $k_z$ is a good quantum number, for each fixed $k_z$, we can consider the model as an effectively 2D system under perpendicular electric and magnetic fields. In such case, it is known that as long as the drift velocity $v_d=E_\text{eff}/B$ is less than the Fermi velocity $v_0$, i.e. when $\beta\equiv w_\bot/v_0<1$, LL solutions exist and can be obtained either by performing a Lorentz boost  to eliminate the $E_\text{eff}$ field~\cite{Luko}, or by using a method from Landau and Lifshitz~\cite{Land,Allan}.

After straightforward but somewhat tedious calculations~\cite{supp}, we find the LL spectrum and the eigenstates for $\beta<1$:
\begin{equation}\label{LLE}
\varepsilon_n(k_z)=w_\| k_z+\text{sgn}(n)\sqrt{\alpha^2v_0^2 k_z^2+|n|\alpha^3\omega_c^2},
\end{equation}
\begin{equation}\label{LLF}
\Psi_n=\frac{1}{\mathcal{N}}e^{ik_x x+ik_z z}e^{-(\text{arctanh}\beta/2)\sigma_x}\left[
                                                                   \begin{array}{c}
                                                                     a_n \phi_{|n|-1} \\
                                                                     -b_n \phi_{|n|} \\
                                                                   \end{array}
                                                                 \right],
\end{equation}
for integers $|n|\geq 1$, where  $\alpha=\sqrt{1-\beta^2}$, $\omega_c=\sqrt{2} v_0/\ell_B$ with $\ell_B=(\frac{1}{eB})^{1/2}$ the magnetic length, $\mathcal{N}$ is a normalization factor, $\phi_m$'s are the harmonic oscillator eigenstates with a scaled $y$-coordinate~\cite{supp}, and the coefficients $a_n=\cos\frac{\zeta}{2}$, $b_n=\sin\frac{\zeta}{2}$ for $n>0$; while $a_n=\sin\frac{\zeta}{2}$, $b_n=-\cos\frac{\zeta}{2}$ for $n<0$, with $\zeta\in[0,\pi]$ satisfying $\tan \zeta =\sqrt{|n|\alpha}\omega_c/(v_0k_z)$. Besides, a Weyl node features an anomalous zeroth LL:
\begin{equation}\label{LL0}
\varepsilon_{n=0}(k_z)=(w_\|-\alpha v_0)k_z,
\end{equation}
with $a_0=0$ and $b_0=1$. The energies in (\ref{LLE}) scale as $\sqrt{B}$ for large $n$ or small $k_z$, which is a characteristic of the linear dispersion.

A key observation is that the cyclotron frequency $\omega_c$, which characterizes the LL spacings, gets reduced by the factor $\alpha(<1)$, arising from the tilt term, to an effective $\omega_c^*=\alpha^{3/2}\omega_c$. Hence, the tilt has the effect of squeezing the LL spectrum. Since $\alpha$ becomes imaginary for $\beta>1$, the LL solution above is valid only for $\beta<1$. For $\bm w$ with a fixed magnitude, the squeezing factor is solely determined by the relative orientations between the tilt and the $B$-field.  By rotating either the sample or the $B$-field, one can continuously tune the degree of LL squeezing.

\begin{figure}[t!]
\includegraphics[width=8.8cm]{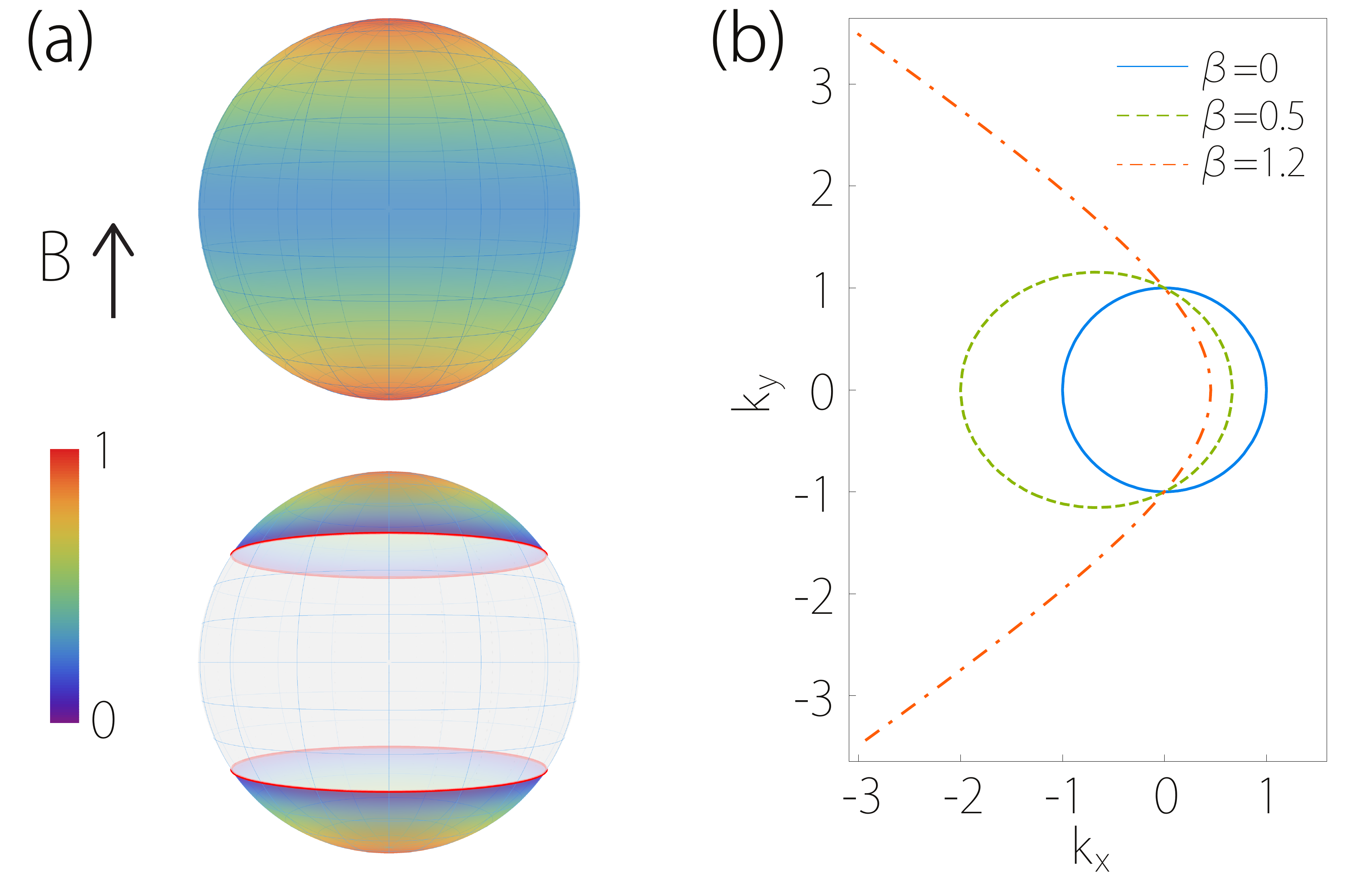}
\caption{(a) LL squeezing factor $\alpha^{3/2}$ plotted versus $\hat{\bm w}$ on a unit sphere. (Upper): conventional Weyl node with $w/v_0=0.9$. (Lower): type-II node with $w/v_0=1.2$, in which the two red loops mark the critical angle where the LLs collapse.
(b) Semiclassical orbit transforms from closed orbit at $\beta<1$ to open trajectory at $\beta>1$. Here $k_z=0$,
$E=0.1$ eV, and the wave-vectors are in units of 0.1 eV/$v_0$.  }
\label{fig1}
\end{figure}

In Fig.~\ref{fig1}(a), we plot the squeezing factor $\alpha^{3/2}$ on a unit sphere denoting the direction of $\hat{\bm w}$. Recall that the conventional Weyl node and the type-II node are distinguished by whether $w/v_0$ is less than 1 or not. Hence for conventional nodes, $\beta<1$ must hold, and the LL solution always exists. The squeezing effect is enhanced when the polar angle $\theta$ of $\hat{\bm w}$ (the angle between $\bm w$ and $\bm B$) increases (decreases) for $w_\|>0$ ($<0$). In contrast, for a type-II node, as $\beta$ can take values larger than 1, there must be a critical angle $\theta_c=\arcsin\frac{v_0}{w}$ for $w_\|>0$ (or $\pi-\arcsin\frac{v_0}{w}$ for $w_\|<0$) beyond which the LL solution in Eqs.(\ref{LLE}-\ref{LL0}) ceases to exist.
Remarkably, approaching the critical angle, $\beta\rightarrow 1$, we have $\alpha,\omega_c^*\rightarrow 0$ and the whole LL spectrum collapses, regardless of the magnetic field strength.

The LL squeezing and collapse can be more easily understood within a semiclassical picture. In semiclassical dynamics, we trace the motion of an electron wave-packet center $(\bm r_c, \bm k_c)$ in both real space and $k$-space. Under a $B$-field, the semiclassical orbit $C$ in $k$-space is residing on the intersection between a constant energy surface and a plane perpendicular to the field direction~\cite{Marder}. It becomes quantized when we apply the Bohr-Sommerfeld quantization condition $\oint_C \bm q_c\cdot d\bm r_c=2\pi[n+\nu/4-\Gamma_C/(2\pi)]$~\cite{Niu,XCN,Cai}, where $\bm q_c=\bm k_c-e \bm A(\bm r_c)$  is the canonical conjugate of $\bm r_c$, $n$ is the quantization integer, $\nu$ is the Maslov index which equals 2 for a closed cyclotron orbit, and $\Gamma_C$ is the Berry phase of the orbit. With the help of the equations of motion, the condition can be expressed as
\begin{equation}\label{semi}
\ell_B^2 A_{C_n}=2\pi\left(n+\frac{1}{2}-\frac{\Gamma_{C_n}}{2\pi}\right),
\end{equation}
where $A_{C_n}$ is the area enclosed by the orbit $C_n$ in $k$-space. Now consider a constant energy surface with energy $E$ for the Hamiltonian (\ref{H0}). As illustrated in Fig.~\ref{fig1}(b), starting from the configuration with $\bm w\| \bm B$, $\beta=0$, the orbit is a circle with a fixed $k_z$. When rotating $\bm w$ away from the $B$-field direction, $\beta$ increases and the orbit for the same $k_z$ and $E$ becomes an ellipse and its area gets increased. According to Eq.(\ref{semi}), the index $n$ associated with the orbit would become larger, which means that more  LLs are squeezed under energy $E$.
A drastic charge occurs when $\beta\rightarrow 1$, during which the area approaches infinity hence the LLs collapse. Beyond this point, as the Weyl cone becomes tipped over in the orbital plane, the semiclassical orbits transforms from closed orbits to open trajectories (hyperbola) (Fig.~\ref{fig1}(b)). Physically, this means that after collapse the dynamics goes beyond the effective model in (\ref{H0}) and a more complete band structure is needed~\cite{supp,Tan}.

From the discussion, it is clear that the collapse depends only on the orientation of the field relative to the tilt but not its strength, and happens only in the type-II case.
In the analysis we did not mention the variation of the Berry phase, because this term is on the order of unity hence does not affect the qualitative conclusion. However, it is indispensable for a quantitative calculation. Particularly, for $k_z=0$, the model is similar to the 2D graphene model, where the $\pi$ Berry phase is crucial for obtaining the correct LL spectrum~\cite{Novo,YBZhang}. Based on Eq.(\ref{semi}), we numerically calculate the LL spectrum for $\beta<1$, which shows excellent agreement with the exact quantum result~\cite{supp}.

Experimentally, the effects of LL squeezing and collapse can be detected, e.g., by scanning tunneling spectroscopy or in Shubnikov-de Haas oscillations. By rotating the sample or the $B$-field, one can find the tilt axis by locating the direction with the least squeezing. The magnitude of the tilt can also be probed by measuring the critical angle.

\begin{figure}[t!]
\includegraphics[width=9.2cm]{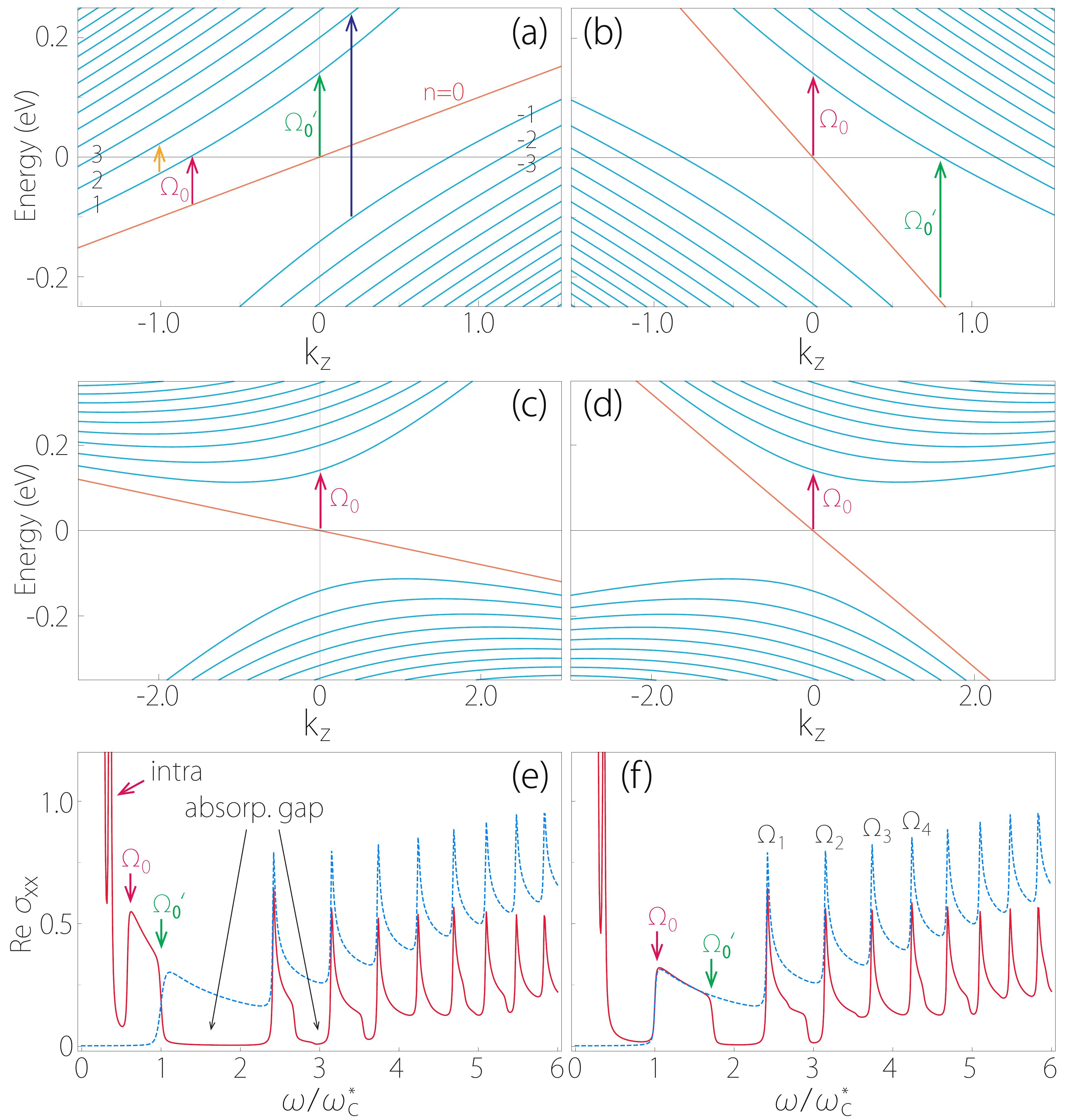}
\caption{LL dispersion along $k_z$ for (a) type-II Weyl node ($w/v_0=2$) and (c) conventional Weyl node ($w/v_0=0.6$), both with $\hat{\bm w}$ along the $B$-field ($w_\|>0$). (b) and (d) are the same as (a) and (c) respectively but with $\hat{\bm w}$ antiparallel to the $B$-field ($w_\|<0$). Arrows in (a) mark some representative optical transitions. (e) Re$\,\sigma_{xx}(\omega)$ plotted for type-II (red solid curve) and conventional Weyl node (blue dashed curve) corresponding to (a) and (c) respectively, in units of $e^2/(2\pi \ell_B)$. (f) is the same as (e) but with a reversed field direction, corresponding to (b) and (d). Here $\beta=0$,  $\omega_c=0.14$ eV, $\mu=0$ eV, and $k_z$ is in units of 0.1 eV/$v_0$. In (e) and (f), $k_BT$ and scattering rate $\Gamma$ are set as $0.01\omega_c$.}
\label{fig2}
\end{figure}

{\color{blue}{\em Optical Conductivity.}}---Magneto-optical measurements provide rich information on the LL structure and electron dynamics. We show that in the regime \emph{before} LL collapse, distinct signatures of type-II nodes would still manifest in the magneto-optical response.
Our focus is on the absorptive part of the longitudinal magneto-optical (\emph{ac}) conductivity, which can be obtained via the Kubo formula:
$
\text{Re}\,\sigma_{xx}(\omega)=-\frac{e^2}{4\pi\ell_B^2}\sum_{nn'}\int dk_z\frac{\Delta f}{\Delta \varepsilon}
|\langle n|\hat v_x|n'\rangle|^2\delta(\omega+\Delta\varepsilon),
$
where $n$ and $n'$ stand for the LL states with the same $k_z$, $\Delta \varepsilon=\varepsilon_n-\varepsilon_{n'}$ and $\Delta f=f(\varepsilon_n)-f(\varepsilon_{n'})$ are the energy and the occupation differences between the two states involved in the optical transition, $f$ is the Fermi-Dirac distribution, and $\hat v_x$ is the velocity operator.

Using the obtained LL solution, we can derive an analytic expression of $\text{Re}\,\sigma_{xx}$~\cite{supp}. The features are most clearly exposed for field directions with small $\beta$ as the LLs are well separated. The result for $\beta=0$ case is plotted in Fig.~\ref{fig2}. Several key observations can be made without resorting to the detailed expression, but by noting that: (1) optical transitions are vertical with conserved $k_z$; (2) at low temperature $T$, transitions are from occupied states to empty states; (3) for $\beta=0$, the optical matrix element $\langle n|\hat v_x|n'\rangle$ is only nonzero for $|n|=|n'|\pm 1$, yielding the familiar selection rule~\cite{Gusy,Tse}.
By inspecting the LL spectra in Fig.~\ref{fig2}(a), we can identify the following distinctive features for type-II nodes.

First, there is invariable presence of intraband absorption peaks at low frequencies (Fig.~\ref{fig2}(e)).
This is because the linear term $w_\| k_z$ for type-II node dominates the LL dispersion in (\ref{LLE}) at large $k_z$.
Considering a positive LL $(n>0)$, at large $k_z$, $\varepsilon_n\sim w_\|k_z+\alpha v_0|k_z|$. Because the condition $|w_\||>|\alpha v_0|$ (in the $\beta<1$ regime)
holds for a type-II node, the energy $\varepsilon_n$ must cross the Fermi level at a negative (positive) $k_z$ for $w_\|>0$ ($<0$), where intraband transitions to its neighboring LLs will occur. Similar conclusion also applies for the negative LLs. In contrast, $|w_\||<|\alpha v_0|$ for conventional Weyl nodes, hence
intraband peaks are absent when $\mu$ is small (Fig.~\ref{fig2}(c,e)) and appear only at higher chemical potentials~\cite{Ashby}.

Secondly, with increasing frequency, interband transition peaks will appear, with distinct shapes. One finds that both $-|n|\rightarrow |n|+1$ and $-(|n|+1)\rightarrow |n|$ transitions have an onset frequency at $\Omega_n=(\sqrt{|n|}+\sqrt{|n|+1})\omega_c^*$ for $|n|\geq 1$. For conventional WSMs or other materials, the peaks typically have long tails because the transitions persist with increasing frequency at larger $k_z$~\cite{Ashby}. In sharp contrast, for a type-II node, due to the above-mentioned unusual $k_z$-dispersion, both positive and negative LLs cross Fermi level at finite $k_z$, hence the allowed transitions between each LL pair are restricted in a finite $k_z$-interval with a finite frequency range, making the peaks tailless. The first few interband peaks can be observed as separated with absorption gaps, strikingly different from that of conventional nodes (Fig.~\ref{fig2}(e)).

This feature is most obvious for the first interband peak involving the zeroth LL. For example, at $\mu=0$, the peaks of $0\rightarrow 1$ and $-1\rightarrow 0$ coincide in the frequency interval $[\Omega_0,\Omega_0']$ with $\Omega_0=\sqrt{\frac{w_\|-\alpha v_0}{w_\|+\alpha v_0}}\omega_c^*$, $\Omega_0'=\omega_c^*$, for $w_\|>0$; whereas $\Omega_0=\omega_c^*$, $\Omega_0'=\sqrt{\frac{|w_\||+\alpha v_0}{|w_\||-\alpha v_0}}\omega_c^*$, for $w_\|<0$.
The difference between positive and negative $w_\|$ originates from the dispersion in (\ref{LL0}) and the condition $|w_\||>|\alpha v_0|$, such that the slope of the zeroth LL must change sign following that of $w_\|$. As a result, the transitions occur at a different $k_z$ interval with a different frequency range (see Fig.~\ref{fig2}(a,b)). In contrast, for conventional nodes, the absorption always starts from the same $\Omega_0$ and has no end frequency when $w_\|$ switches sign (Fig.~\ref{fig2}(c,d)).

Thirdly, from the above discussion, distinct signatures appear when varying the $B$-field direction. Most interestingly, when reversing the $B$-field direction, which is equivalent to switching the sign of $w_\|$, all the interband peak positions $\Omega_n$ remain unchanged except for $\Omega_0$, as shown in Fig.~\ref{fig2}(e,f). As discussed, this effect stems from the unusual $k_z$-dispersion of the zeroth LL and is unique to the type-II Weyl node. Due to the squeezing factor in $\omega_c^*$, the peak positions can be continuously tuned by rotating the sample or the $B$-field, and the peaks are squeezed to the low-frequency end when approaching the critical angle of LL collapse.
Therefore by tracking the absorption peaks, we could distinguish type-II nodes and further extract information of the tilt.

When $\mu$ is tuned away from the node, the peaks of $0\rightarrow 1$ and $-1\rightarrow 0$ will begin to split. And the frequency $\Omega_n$ will be shifted once $\mu$ passes the LL energy $\varepsilon_{|n+1|}$ at $k_z=0$.
For $\beta\neq 0$, additional $n\rightarrow m$ transitions become possible~\cite{Sari}, leading to additional absorption peaks that scale as $\beta^2$ for small $\beta$~\cite{supp}.
Finite temperatures and disorder scattering both smooth out the absorption profile. The scattering effects may be captured phenomenologically by broadening the delta function in the Kubo formula to a Lorentzian with a width $\Gamma$ representing the scattering rate  (the \emph{dc} limit $\sigma_{xx}(\omega=0)$ diverges as $1/\Gamma$, as in Drude model). The key features in the absorption spectrum would be observable as long as $k_B T$ and $\Gamma$ are small compared with $\omega_c^*$.

{\color{blue}{\em Discussion.}}---The effect of LL collapse has previously been discussed in the context of 2D Dirac systems~\cite{Luko,Goerbig,Goerbig2,Gu,Peres,Wei,Sari,Pros}. There, the electron motion is confined within the 2D plane, the collapse requires a typically large in-plane $E$-field and also depends on the strength of the $B$-field, making such experiment quite a challenge. However, for type-II WSMs, the collapse does not require any external $E$-field, and is independent of the $B$-field strength, which should facilitate its experimental realization. Being a 3D system, the orbital plane rotates as the field direction varies in space, continuously changing the LL spectrum. The identified features in the optical absorption are tied with the special LL dispersion along the field, hence are unique for 3D systems with no analog in 2D. Moreover, the features are most obvious for those involving the zeroth LL which is unique for Weyl nodes. Therefore they indeed constitute unique signatures for type-II Weyl nodes, distinct from conventional WSMs and other materials.

As mentioned, in a WSM, Weyl nodes always occur in pairs of opposite chirality. Additionally, a WSM phase cannot exist if both time reversal $(\mathcal{T})$ and inversion $(\mathcal{P})$ symmetries are present. In the simplest case with broken $\mathcal{T}$, a WSM can have a single pair of Weyl nodes: the partner of the node in (\ref{H0}) will have opposite chirality and a reversed tilt vector, if $\mathcal{P}$ is preserved. The magneto-response studied here is identical for the two nodes. On the other hand, if $\mathcal{P}$ is broken, the two nodes related by $\mathcal{T}$ are of the same chirality while $\bm w$ is reversed, and the magneto-response of the partner is effectively the same as (\ref{H0}) but with a reversed $B$-field~\cite{supp}. In the presence of multiple pairs of nodes, the magneto-response are generally different for each one, unless tied by symmetry. One can expect interesting cases such as different onsets of LL collapse at different nodes when rotating the $B$-field.

We used an isotropic Fermi velocity in the analysis. Generally, the Fermi velocity can be different along the three principal axes, which, however, does not affect the main conclusions regarding the LL collapse and the key features in magneto-optical response. In fact, one can  rescale the coordinates to map such case to model (\ref{H0})~\cite{supp}.

Finally, in a type-II WSM, the type-II nodes occur in-between electron and hole pockets~\cite{Been}, and  other conventional bands may also appear around the Fermi level. However their magneto-responses are different, such as the different scaling of LL spacings ($\propto B$) and the absence of LL collapse, hence the signals from the type-II node should still be detectable in experiment.

\begin{acknowledgements}


{\color{blue}{\em Note added.}}---Recently, two complementary and independent studies~\cite{Udagawa,Tchoumakov} appeared, with  a similar topic via different approaches.
\end{acknowledgements}

\bibliographystyle{apsrev4-1}

\end{document}


\title{Supplemental Material for ``Unusual Magneto-Response in Type-II Weyl Semimetals"}

\author{Zhi-Ming Yu}
\affiliation{Beijing Key Laboratory of Nanophotonics and Ultrafine Optoelectronic Systems, School of Physics, Beijing Institute of Technology, Beijing 100081, China}

\author{Yugui Yao}
\affiliation{Beijing Key Laboratory of Nanophotonics and Ultrafine Optoelectronic Systems, School of Physics, Beijing Institute of Technology, Beijing 100081, China}

\author{Shengyuan A. Yang}
\affiliation{Research Laboratory for Quantum Materials, Singapore University of Technology and Design, Singapore 487372, Singapore}


\begin{abstract}
\end{abstract}


\maketitle
\section{Solution of LLs by Lorentz boost\label{sub:LB}}
A simple $2\times2$ Weyl Hamiltonian for the low-energy states near
a Weyl node with tilted spectrum reads:
\begin{eqnarray}
H & = & v_{0}\boldsymbol{k\cdot\sigma}+\boldsymbol{w\cdot k}.\label{eq:hampri}
\end{eqnarray}
As discussed in the main text, without loss of generality, we could choose our coordinates, such that the $z$-axis is along the $B$-field and the tilt vector is in the $x$-$z$ plane with the form $\boldsymbol{w}=(\begin{array}{ccc} w_{\bot}, & 0, & w_{\|}\end{array})$.
Under a magnetic field ($\boldsymbol{B}=B\hat{{z}}$),
we make the usual Peierls substitution $\boldsymbol{k}\rightarrow\boldsymbol{k}+e\boldsymbol{A}$
($e>0$ and we set $\hbar=1$ here) and choose the Landau gauge ($\boldsymbol{A}=-By\hat{{x}}$) in the Hamiltonian.
Since $k_{x}$ and $k_{z}$ are good quantum number, we can solve the spectrum by introducing an alternative Hamiltonian:
\begin{eqnarray}
{\cal H}_{L} & = & {\cal H}-w_{\bot}k_{x}-w_{\|}k_{z}=v_{0}\left(\boldsymbol{k}+e\boldsymbol{A}\right)\boldsymbol{\cdot\sigma}-eByw_{\bot},
\end{eqnarray}
which is reminiscent of an effective 2D Dirac system with a mass term $v_{0}k_{z}\sigma_{z}$ under a magnetic field ($\boldsymbol{B}=B\hat{{z}}$)
and an effective electric field ($\boldsymbol{E}_\text{eff}=-w_{x}B\hat{{y}}$). When $\beta=w_\bot/v_0<1$,
the effective electric field can be eliminated by a Lorentz boost\cite{Lukose}.
To demonstrate this, it is convenient to rewrite the equation in a manifestly covariant time-dependent form:
\begin{eqnarray}
i\gamma_{\mu}\left(\partial^{\mu}+ieA^{\mu}\right)\Psi(x^{\mu}) & = & k_{z}\Psi(x^{\mu})\label{eq:coveq1}
\end{eqnarray}
where $x^{\mu}=(\begin{array}{ccc}
v_{0}t, & x, & y\end{array})$, $\gamma_{\mu}=(\begin{array}{ccc}
\sigma_{z}, & \sigma_{z}\sigma_{x}, & \sigma_{z}\sigma_{y}\end{array})$ and $A^{\mu}=(\begin{array}{ccc}
A_{x}w_{\bot}/v_{0}, & A_{x}, & 0\end{array})$.
Applying a Lorentz boost in the $x$-direction,
\begin{eqnarray}
\left(\begin{array}{c}
\tilde{x}^{0}\\
\tilde{x}^{1}
\end{array}\right) & = & \left(\begin{array}{cc}
\cosh U & \sinh U \\
\sinh U & \cosh U
\end{array}\right)\left(\begin{array}{c}
x^{0}\\
x^{1}
\end{array}\right),
\end{eqnarray}
with $\tilde{x}^{2}=x^{2}$ and choosing
$\tanh U\equiv \beta=A^{0}/A^{1}=w_{\bot}/v_{0}$,
Eq.~(\ref{eq:coveq1}) can be rewritten as:
\begin{eqnarray}
i\gamma_{\mu}\left(\tilde{\partial}^{\mu}+ie\tilde{A}^{\mu}\right)\tilde{\Psi}(\tilde{x}^{\mu}) & = & k_{z}\tilde{\Psi}(\tilde{x}^{\mu}),\label{eq:coveq2}
\end{eqnarray}
where $\tilde{A}^{\mu}=(\begin{array}{ccc}
0, & A_{x}\alpha, & 0\end{array})$ with $\alpha\equiv\sqrt{1-\beta^{2}}$ and $\tilde{\Psi}(\tilde{x}^{u})=e^{\sigma_{x}U/2}\Psi(x^{\mu})$.
The above equation (\ref{eq:coveq2}) is nothing but the equation of a gapped 2D Dirac system under a reduced magnetic
field $\tilde{\boldsymbol{B}}=\alpha B\hat{{z}}$.
The eigenvalues and eigenfunctions of (\ref{eq:coveq2}) are given by
\begin{eqnarray}
\tilde{\varepsilon}_{L,n} & = & \text{sgn}\left(n\right) v_{0}\frac{1}{\tilde{\ell}_{B}}\sqrt{2|n|+k_{z}^{2}\tilde{\ell}_{B}^{2}},\\
\tilde{\Psi}_{n} & \propto & e^{ik_{z}z+i\tilde{k}_{x}\tilde{x}}\left[\begin{array}{c}
a_{n}\phi_{|n|-1}(\xi)\\
-b_{n}\phi_{|n|}(\xi)
\end{array}\right],
\end{eqnarray}
for $|n|\geq 1$,  where $\tilde{\ell}_{B}=\sqrt{1/(e\alpha B)}$ is the re-scaled
magnetic length, $\phi_{|n|}$ are the harmonic oscillator eigenstates, with $\xi  =  (\tilde{y}-\tilde{\ell}_{B}^{2}\tilde{k}_{x})/\tilde{\ell}_{B}$
and the coefficients $(\begin{array}{cc}
a_{n}, & b_{n}\end{array})=(\begin{array}{cc}
\cos\frac{\zeta}{2}, & \sin\frac{\zeta}{2}\end{array})$ for $n>0$ while $(\begin{array}{cc}
a_{n}, & b_{n}\end{array})=(\begin{array}{cc}
\sin\frac{\zeta}{2}, & -\cos\frac{\zeta}{2}\end{array})$ for $n<0$ with $\zeta\in\left[0,\ \pi\right]$ satisfying $\tan\zeta=\sqrt{|n|}\tilde{\omega}_c/(v_0k_{z})$ with $\tilde{\omega}_c=\sqrt{2}v_0/\tilde{\ell}_{B} $.
In addition, there is an additional zeroth LL:
\begin{eqnarray}
\tilde{\varepsilon}_{L,n=0} & = & -v_{0}k_{z},
\end{eqnarray}
with $(\begin{array}{cc}
a_{0}, & b_{0}\end{array})=(\begin{array}{cc}
0, & 1\end{array})$. Applying the inverse boost transformation, the physical energy eigenvalues of the original problem are obtained:
\begin{eqnarray}
\varepsilon_{n} & = & \varepsilon_{L,n}+w_{\bot}k_{x}+w_{\|}k_{z}=\begin{cases}
w_{\|}k_{z}+\text{sgn}\left(n\right)\sqrt{\alpha^{2}v_{0}^{2}k_{z}^{2}+|n|\alpha^{3}\omega_{c}^{2}} & |n|\geq 1;\\
\left(w_{\|}-\alpha v_{0}\right)k_{z} & n=0,
\end{cases}\label{eq:LBv1}
\end{eqnarray}
with $\omega_{c}=\sqrt{2}v_{0}/\ell_{B}$ and $\ell_{B}=\sqrt{1/(eB)}$.
The corresponding eigenfunctions are
\begin{eqnarray}
\Psi_{n} & \propto & e^{ik_{z}z+ik_{x}x}e^{-\sigma_{x}U/2}\left[\begin{array}{c}
a_{n}\phi_{|n|-1}(\xi)\\
-b_{n}\phi_{|n|}(\xi)
\end{array}\right],\label{eq:LBv2}
\end{eqnarray}
with a scaled $y$-coordinate
$\xi = \sqrt{\alpha}\left(y-\ell_{B}^{2}k_{x}\right)/\ell_{B}-\text{sgn}(n)\beta\sqrt{2|n|+k_{z}^{2}\ell_{B}^{2}\alpha^{-1}}$.

\section{LL spectrum from semiclasscal quantization}
\begin{figure}[h!]
\includegraphics[width=11cm]{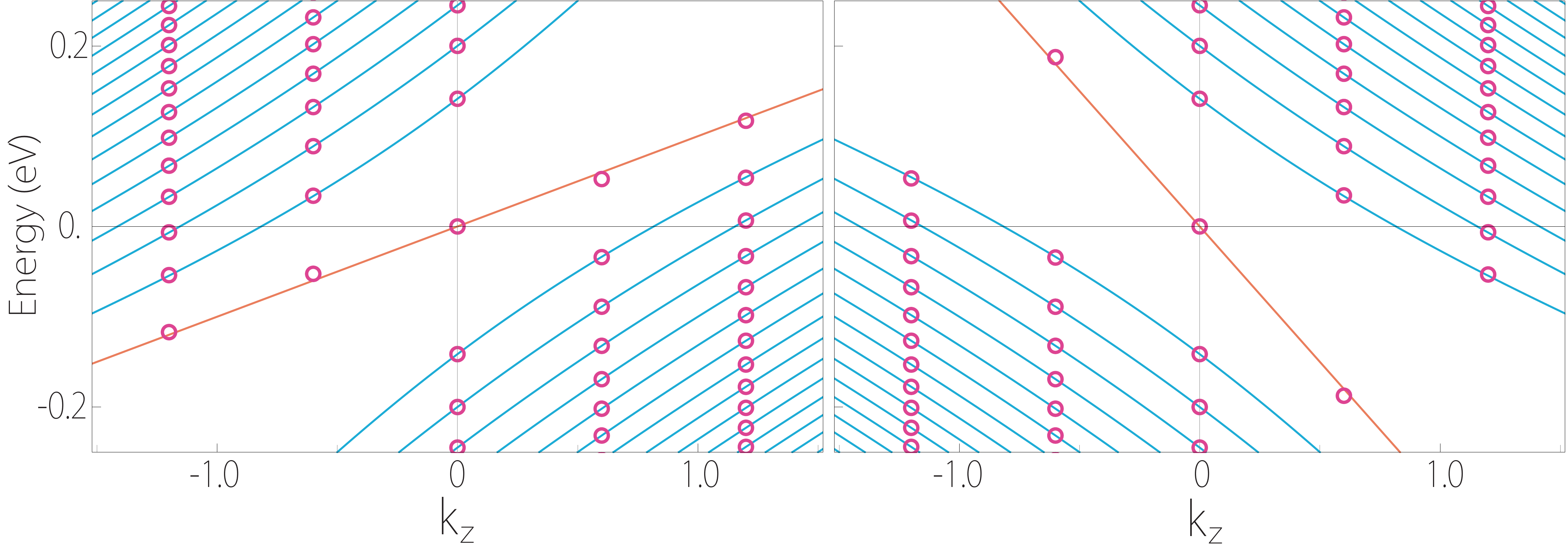}
\caption{LL spectrum for type-II Weyl node with (left)  $\hat{\bm w}$ along the $B$-field ($w_\|>0$) and (right) $\hat{\bm w}$ antiparallel to the $B$-field ($w_\|<0$). The red circles are obtained from the semiclassical quantization condition, and the solid lines are the exact quantum result. The parameters used here are the same as those for Fig.2(a,b) in the main text.}
\label{figs1}
\end{figure}

As discussed in the main text, the Bohr-Sommerfeld quantization condition combined with the semiclassical dynamics leads to the following equation~\cite{Niu,XCN,Cai}
\begin{equation}\label{semi}
\ell_B^2 A_{C_n}=2\pi\left(n+\frac{1}{2}-\frac{\Gamma_{C_n}}{2\pi}\right).
\end{equation}
To numerically evaluate the LLs, we obtain the band energy in the absence of $B$-field,
\begin{equation}
\varepsilon(\bm k)=w_x k_x+w_z k_z\pm v_0 k.
\end{equation}
The $B$-field along $z$-axis shifts the wave-packet energy. To linear order, the corrected wave-packet energy is given by~\cite{Cai}
\begin{equation}
\mathcal E(\bm k)=\varepsilon(\bm k)-M_z(\bm k)B,
\end{equation}
where $M_z(\bm k)=-ev_0 k_z/(2k^2)$
is the ($z$-component of the) orbital magnetic moment. The semiclassical orbit is at the intersection between a surface with constant energy $\mathcal E(\bm k)$ and a plane with constant $k_z$. The Berry phase in (\ref{semi}) can be evaluated using
\begin{equation}
\Gamma_{C_n}=\int_{S_{C_n}} \Omega_z(\bm k)d^2 k,
\end{equation}
where $\Omega_z(\bm k)=-k_z/(2 k^3)$
is the ($z$-component of the) Berry curvature. The numerical result of the semiclassical LLs is plotted in Fig.~\ref{figs1} to compare with the exact quantum result. One indeed observes a very good agreement.

\section{Magneto-optical conductivity}
The longitudinal magneto-optical conductivity $\sigma_{xx}$ is given by
\begin{eqnarray}
\sigma_{xx}(\omega) & = & -\frac{ie^{2}}{2\pi\ell_{B}^{2}}\sum_{nn^{\prime}}\int\frac{dk_{z}}{2\pi}\frac{f(\epsilon_{n})-f(\epsilon_{n^{\prime}})}{\varepsilon_{n}-\varepsilon_{n^{\prime}}}\times\frac{\langle n|v_{x}|n^{\prime}\rangle\langle n^{\prime}|v_{x}|n\rangle}{\omega+\varepsilon_{n}-\varepsilon_{n^{\prime}}+i0^+},
\end{eqnarray}
where $|n\rangle\equiv|\psi\rangle_{n}$ is the LL state with a fixed $k_z$, and $f(x)=1/(1+e^{\beta(x-\mu)})$ is the Fermi-Dirac distribution,
$\beta$ is the inverse temperature, and $\mu$ is the chemical potential.
The velocity operator is given by
\begin{eqnarray}
v_{x} & = & \frac{\partial{\cal H}}{\partial k_{x}}=w_{x}+v_{0}\sigma_{x}.
\end{eqnarray}
The dissipative component of the longitudinal conductivity is
\begin{eqnarray}
\text{Re}(\sigma_{xx}) & = & -\frac{e^{2}}{4\pi\ell_{B}^{2}}\sum_{mn}\int dk_{z}\frac{f(\epsilon_{m})-f(\epsilon_{n})}{\varepsilon_{m}-\varepsilon_{n}}|\langle m|v_{x}|n\rangle|^{2}\delta\left(\omega+\varepsilon_{m}-\varepsilon_{n}\right).
\end{eqnarray}
Disorder scattering effects can be included by broadening the delta function to a Lorentzian with a width $\Gamma$ representing the scattering rate.
{ After straightforward but somewhat tedious calculation, we find that the optical matrix element can be expressed as
\begin{eqnarray}
|\langle m|v_{x}|n\rangle|^{2} & = & v_{x,0}^{mn}+v_{x,1}^{mn}\cos^{2}\phi
\end{eqnarray}
where $\phi$ is the azimuthal angle of $\boldsymbol{w}$, leading to the $\phi$ dependence of the strength of the absorption peaks and
\begin{eqnarray*}
v_{x,0}^{mn}=\frac{v_{0}^{2}\alpha^{2}\left({\cal Q}_{1}-{\cal Q}_{2}\right)^{2}}{\sqrt{\left(1+g_{m}^{2}\right)\times\left(1+g_{n}^{2}\right)}} & ; & v_{x,1}^{mn}=\frac{v_{0}^{2}\alpha^{2}\left[4{\cal Q}_{1}{\cal Q}_{2}-\beta^{2}\left({\cal Q}_{1}+{\cal Q}_{2}\right)^{2}\right]}{\sqrt{\left(1+g_{m}^{2}\right)\times\left(1+g_{n}^{2}\right)}}
\end{eqnarray*}
where $g_{n}(k_{z})\equiv\left(\text{sgn}\left(n\right)\sqrt{v_{0}^{2}k_{z}^{2}+|n|\alpha\omega_{c}^{2}}-v_{0}k_{z}\right)/\left(\sqrt{\alpha|n|}\omega_{c}\right)$ , ${\cal Q}_{1}\equiv|C_{|m|,|n|-1}^{mn}|g_{m}$, ${\cal Q}_{2}\equiv|C_{|m|-1,|n|}^{mn}|g_{n}$,
and
\begin{eqnarray}
C_{m'n'}^{mn}  & = & (-1)^{m'}\frac{e^{-|Z_{nm}|^{2}/2}}{\sqrt{m'!n'!}}\left(Z_{mn}^{*}\right)^{n'-m'}\mathcal{U}(-m',1-m'+n',|Z_{nm}|^{2}),
\end{eqnarray}
with $Z_{nm}\equiv  \frac{\beta(\varepsilon_{n}-\varepsilon_{m})}{\omega_{c}\alpha^{3/2}}e^{-i\phi}$ and $\mathcal{U}(a,b,Z)$ is the tricomi confluent hypergeometric function.
From the expression of $C_{m'n'}^{mn}$, one can find that:
(1) For the case of $\beta=0$ (field parallel or antiparallel to the tilt direction), one  has $Z_{mn}=0$, and $C_{m'n'}^{mn}$ is nonzero only when
$m'=n'$, thus the matrix element  $|\langle m|v_{x}|n\rangle|^{2}\sim \delta_{|m|,|n|-1}g_{m}^{2}+\delta_{|m|-1,|n|}g_{n}^{2}$
corresponding to the familiar selection rules $|n|\rightarrow|n|\pm1$ as in graphene.
(2) For $\beta\neq0$, there are  additional non-dipolar  transitions.
For small $\beta$, the strength of these additional transition is proportional to $\beta^2$.
They lead to additional absorption peaks, as shown in Fig.~\ref{figs2}, but do not change the main features discussed in the main text.

\begin{figure}[h!]
\includegraphics[width=11cm]{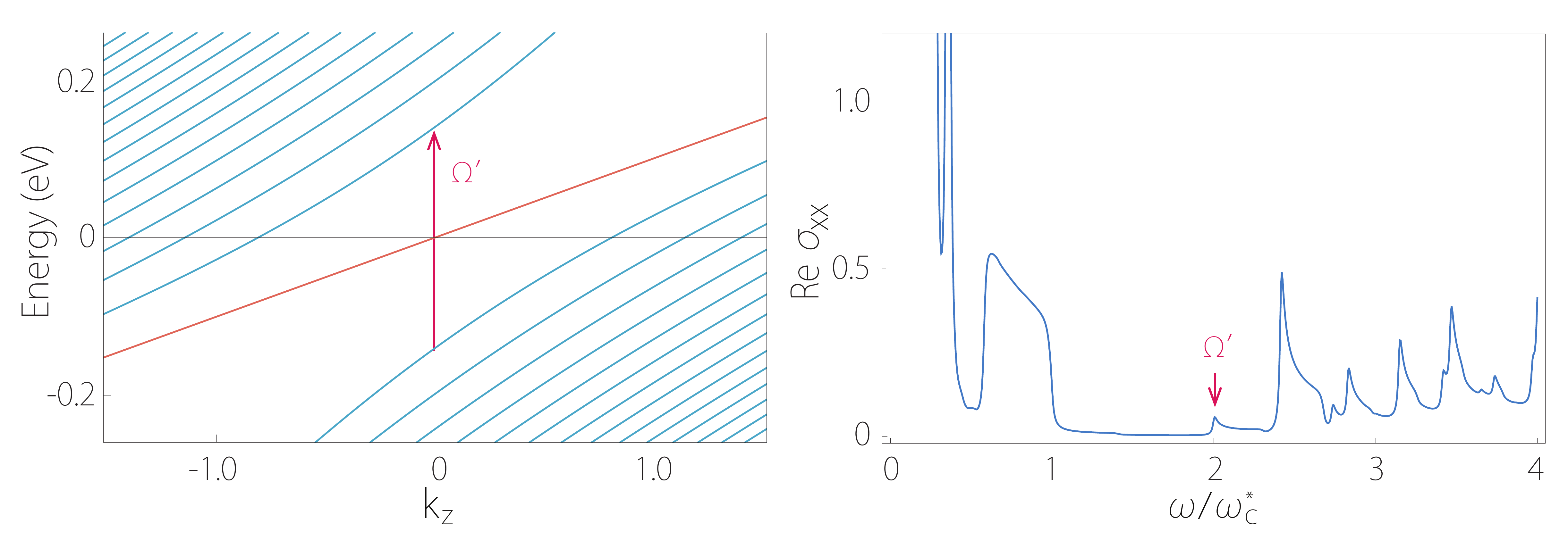}
\caption{Left: LL dispersion along $k_z$ with $w/v_0=2$ and $\beta=0.1$ ($w_\|>0$). The arrow marks the non-dipolar $-1\rightarrow 1$ transition.
Right: Re$\,\sigma_{xx}(\omega)$ plotted (in units of $e^2/(2\pi \ell_B)$) corresponding to left figure. The additional peaks corresponding to the $-1\rightarrow 1$ transition is indicated. Here $\mu=0$, and we choose $\omega_c=0.14$ eV, and $k_BT$ and $\Gamma$ are set as $0.01\omega_c^*$.}
\label{figs2}
\end{figure}

\begin{figure}[h!]
\includegraphics[width=11cm]{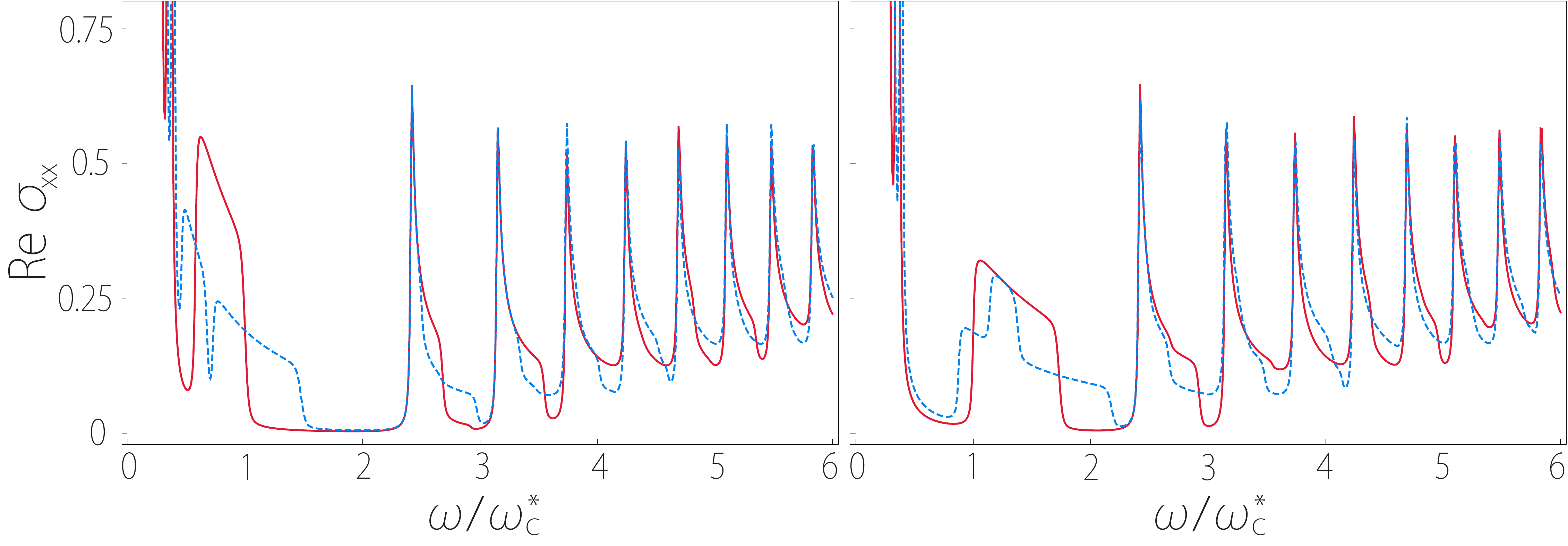}
\caption{Re$\,\sigma_{xx}(\omega)$ plotted (in units of $e^2/(2\pi \ell_B)$) for $\mu=0.4 \omega_c$ (red solid curve) and $\mu=0$ (blue dashed curve) of a type-II Weyl node. Left: $\hat{\bm w}$ is along the $B$-field ($w_\|>0$). Right: $\hat{\bm w}$ is antiparallel to the $B$-field ($w_\|<0$). In the figures, $\beta=0$, and we choose $w/v_0=2$, $\omega_c=0.14$ eV, and $k_BT$ and $\Gamma$ are set as $0.01\omega_c^*$.}
\label{figs3}
\end{figure}

The expression for $\text{Re}(\sigma_{xx})$ is much simplified for the $\beta=0$ case:
\begin{eqnarray}
\text{Re}(\sigma_{xx}) & = & -\frac{e^{2}}{4\pi \ell_{B}^{2}}\left(\Lambda+{\cal V}_{1;k_{z1}}+{\cal V}_{-1;-k_{z1}}\right)
\end{eqnarray}
with $k_{z1} =  (\omega^{2}-\omega_{c}^{2}\alpha^{3})/(2\alpha\omega)$ and
\begin{eqnarray}
{\cal V}_{\lambda;k_{z}} & = & \frac{\alpha\left(\varepsilon_{1}-w_{\|}k_{z}\right)\left[f(\epsilon_{0})-f(\epsilon_{\lambda})\right]}{\left(1+g_{\lambda}^{2}\right)\left(\epsilon_{0}-\epsilon_{\lambda}\right)\times|\varepsilon_{1}-\left(w_{z}-\lambda v_{0}\alpha\right)k_{z}|},\nonumber\\
\Lambda & = & \sum_{n=1}^{n_{max}}\sum_{\lambda=\pm1}\left[\left({\cal {\cal L}}_{n+1;\lambda k_{z2}}^{n}+{\cal {\cal L}}_{-(n+1);\lambda k_{z2}}^{-n}\right)\Theta\left(\omega_{c}^{*}-\omega\right)+\left({\cal {\cal L}}_{n+1;\lambda k_{z2}}^{-n}+{\cal {\cal L}}_{-(n+1);\lambda k_{z2}}^{n}\right)\Theta\left(\omega-\omega_{c}^{*}\right)\right]
\end{eqnarray}
where $\omega_{c}^{*}=\omega_{c}\alpha^{3/2}$,   $k_{z2}=\sqrt{\omega_{c}^{4}\alpha^{6}-2(1+2|n|)\omega_{c}^{2}\alpha^{3}\omega^{2}+\omega^{4}}/(2\alpha\omega)$,  $\Theta(x)$ is the Heaviside step function, $n_{max}$ is the maximum positive  integer, with which $k_{z2}$ still is real, and
\begin{eqnarray}
{\cal {\cal L}}_{n;k_{z}}^{n^{\prime}} & = & \frac{|\varepsilon_{n^{\prime}}-w_{\parallel}k_{z}|\times |\varepsilon_{n}-w_{\parallel}k_{z}|}{|k_{z}\Delta\varepsilon_{nn^{\prime}}|}\times\frac{g_{n^{\prime}}^{2}\Delta f_{nn^{\prime}}}{\alpha v_{0}\left(1+g_{n}^{2}\right)\left(1+g_{n^{\prime}}^{2}\right)\Delta\varepsilon_{nn^{\prime}}},
\end{eqnarray}
with $\Delta f_{nn^{\prime}}=f(\varepsilon_{n})-f(\varepsilon_{n^{\prime}})$,
$\Delta\varepsilon_{nn^{\prime}}=\varepsilon_{n}-\varepsilon_{n^{\prime}}$.

At $\mu=0$, the contributions from the transition channels $-n\rightarrow n+1$ and $-n-1\rightarrow n$ to $\text{Re}(\sigma_{xx})$ are the same.
While for the case of  $\mu\neq0$, these two contribution will generally be different. This will change the shape of the absorption peaks. As $\mu$ deviates from zero, the most affected peak is the first interband peak which involves the zeroth LL.
In Fig.~\ref{figs3}, we compare the results for $\mu=0$ and $\mu =0.4\omega_c$. The major difference is that the two peaks of $0\rightarrow 1$ and $-1\rightarrow 0$ transitions are now split. Since $\mu$ is still less than $\varepsilon_1$, the onset frequencies $\Omega_n$ for higher interband peaks are not affected, despite slight changes at the peak shoulders.

\section{A pair of Weyl nodes}

In a WSM, Weyl nodes always occur in pairs of opposite chirality.
Additionally, the WSM phase cannot exist if both time reversal (${\cal T}$)
and inversion (${\cal P}$) symmetries are present. Here, we consider the
two simplest cases: (a) ${\cal T}$ is broken while ${\cal P}$ is preserved;
and (b) ${\cal P}$ is broken while ${\cal T}$ is preserved.

Case (a): In this case, there could be a single pair of Weyl nodes in the Brillouin zone. For one node located at $\bm K$ described by our model
\begin{eqnarray}\label{46}
H & = & v_{0}\boldsymbol{k\cdot\sigma}+\boldsymbol{w\cdot k},
\end{eqnarray}
where $\bm k$ is measured from $\bm K$, its partner located at $-\bm K$ and described by
\begin{eqnarray}
H_{{\cal P}} & = & -v_{0}\boldsymbol{k\cdot\sigma}-\boldsymbol{w\cdot k}.
\end{eqnarray}
due to the inversion symmetry.
These two Weyl nodes have opposite chirality as well as the tilt vector. The magneto-optical properties of them are found to be identical, as shown in Fig.~\ref{figs4}(a).

Case (b): In this case, for one node located at $\bm K$ described by our model
\begin{eqnarray}
H & = & v_{0}\boldsymbol{k\cdot\sigma}+\boldsymbol{w\cdot k},
\end{eqnarray}
there is another $\mathcal{T}$-related node located at $-\bm K$, described by
\begin{eqnarray}\label{49}
H_{{\cal T}} & = & -v_{0}k_{x}\sigma_{x}+v_{0}k_{y}\sigma_{y}-v_{0}k_{z}\sigma_{z}-\boldsymbol{w\cdot k}.
\end{eqnarray}
These two nodes are of the same chirality while the tilt vectors are opposite.
Under a canonical transformation $\sigma_{y}$, we have
\begin{eqnarray}\label{50}
H_{{\cal T}}^{\prime} & = & \sigma_{y}H_{{\cal T}}\sigma_{y}=v_{0}\boldsymbol{k\cdot\sigma}-\boldsymbol{w\cdot k}.
\end{eqnarray}
Therefore the magneto-response of $H_{\cal{T}}$  is effectively the same as $H$ but with a reversed B-field, as shown in Fig.~\ref{figs4}(b). In deriving (\ref{49}), we assumed that the $\sigma$'s represent certain orbital degree of freedom not reversed under $\mathcal{T}$. If they correspond to a spin degree of freedom, then we would directly obtain the model in Eq.(\ref{50}).

Since the two nodes are of the same chirality, there must exist at least another two nodes with the opposite chirality. Their magneto-response would generally be different from these two, unless additional crystalline symmetries are present to connect these nodes.

\begin{figure}[h!]
\includegraphics[width=11cm]{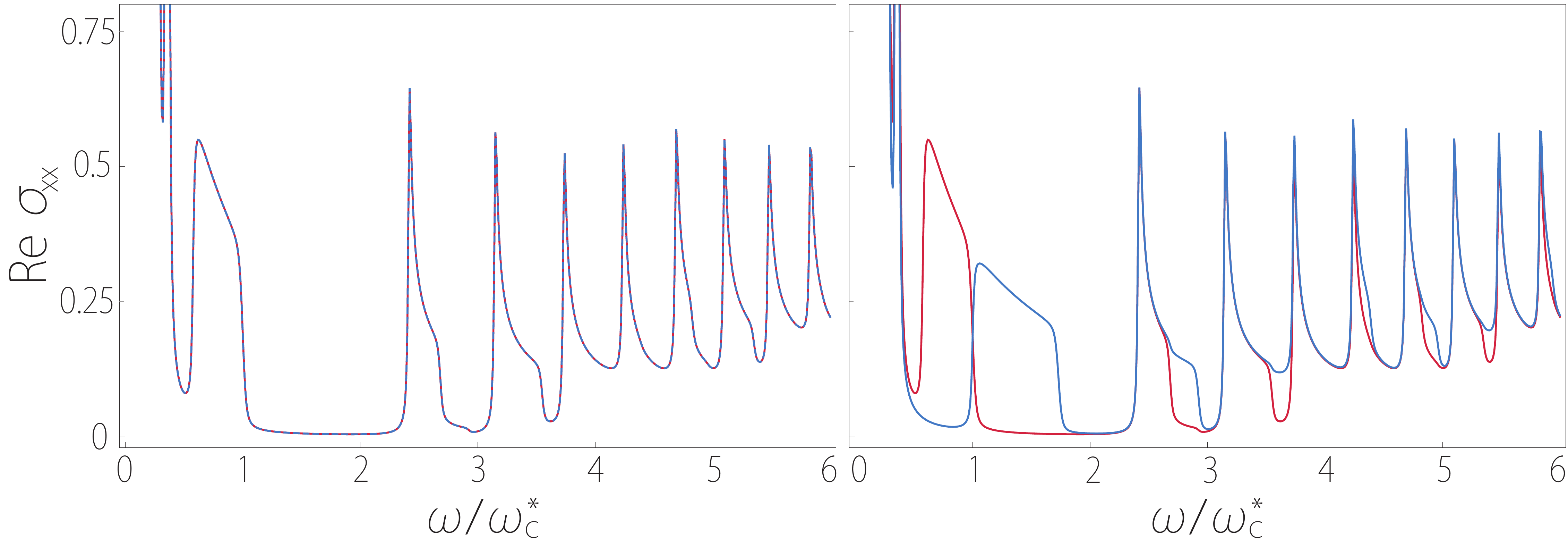}
\caption{Re $\sigma_{xx}(\omega)$ (in units of $e^2/(2\pi \ell_B)$) for a pair of type-II Weyl nodes connected by symmetry. The red curve represents the response from the node described by model (\ref{46}), while the blue curve is from the partner node related by (left figure) inversion symmetry or (right figure) time reversal symmetry. In the left figure, the response from the two nodes are identical, so the two curves overlap.
In the figures, $\hat{\bm w}$ is along the $B$-field, hence $\beta=0$. And we choose $w/v_0=2$, $\omega_c=0.14$ eV, and $k_BT$ and $\Gamma$ are set as $0.01\omega_c^*$.}
\label{figs4}
\end{figure}

\section{Model with anisotropic Fermi velocity}
Let's consider the following model with anisotropic Fermi velocities
\begin{eqnarray}
H & = & \boldsymbol{w\cdot k}+\sum_{i}v_{i}k_{i}\sigma_{i}.
\end{eqnarray}
Here the Fermi velocities $v_i$'s may be different if there is no crystalline symmetry connecting them. The corresponding Hamiltonian under a magnetic field reads:
\begin{eqnarray}
H & = & \boldsymbol{w\cdot}(\boldsymbol{k}+e\boldsymbol{A})+\sum_{i}v_{i}\left(k_{i}+eA_{i}\right)\sigma_{i}
 = \sum_{i}\left[\frac{w_{i}}{\lambda_{i}}\lambda_{i}\left(k_{i}+eA_{i}\right)+v_{0}\lambda_{i}\left(k_{i}+eA_{i}\right)\sigma_{i}\right],\label{eq:HamAni}
\end{eqnarray}
where $v_{0}\equiv(v_{x}v_{y}v_{z})^{1/3}$, $\lambda_{i}\equiv v_{i}/v_{0}$
and $\boldsymbol{A}$ is the vector potential  satisfying $\boldsymbol{B}=\boldsymbol{\nabla}\times\boldsymbol{A}$.
We can map the original problem of an anisotropic model (without the
tilt term) to a problem for an isotropic model by re-scaling the three
axis $k_{i}^{\prime}=\lambda_{i}k_{i}$ and  $r_{i}^{\prime}=r_{i}/\lambda_{i}$,
then  the Hamiltonian  (\ref{eq:HamAni}) can be written as:
\begin{eqnarray}
H & = & \boldsymbol{w}_{i}^{\prime}\boldsymbol{\cdot}\left(\boldsymbol{k}_{i}^{\prime}+e\boldsymbol{A}_{i}^{\prime}\right)+v_{0}\left(\boldsymbol{k}_{i}^{\prime}+e\boldsymbol{A}_{i}^{\prime}\right)\boldsymbol{\cdot}\boldsymbol{\sigma}_{i}
\end{eqnarray}
with $w_{i}^{\prime}=w_{i}/\lambda_{i}$, $A_{i}^{\prime}=\lambda_{i}A_{i}$, and with the $B$-field given by
$ B_{i}^{\prime} = B_{i}/\lambda_{i}$.

For this isotropic model, we could make a rotation of the $r'$-coordinate
system such that the $B$-field is along the new $z$-axis and the tilt
vector is in the new $x$-$z$ plane. Let $\theta_{B^{\prime}}$ and
$\phi_{B^{\prime}}$ be the spherical angles of $B'$ in $r'$-system.
\begin{eqnarray}
\boldsymbol{r}'' & = & R_{y}(\theta_{B'})R_{z}(\phi_{B'})\boldsymbol{r}',
\end{eqnarray}
with
\begin{eqnarray}
R_{y}(\theta_{B'})=\left(\begin{array}{ccc}
\cos\theta_{B'} & 0 & -\sin\theta_{B'}\\
0 & 1 & 0\\
\sin\theta_{B'} & 0 & \cos\theta_{B'}
\end{array}\right) & ; & R_{z}(\phi_{B'})=\left(\begin{array}{ccc}
\cos\phi_{B'} & \sin\phi_{B'} & 0\\
-\sin\phi_{B'} & \cos\phi_{B'} & 0\\
0 & 0 & 1
\end{array}\right).
\end{eqnarray}
In $k$-space, the rotation is the same $\boldsymbol{k}'' =  R_{y}(\theta_{B'})R_{z}(\phi_{B'})\boldsymbol{k}'$.
One checks that $\boldsymbol{B}'' =  \boldsymbol{B}'R_{z}^{-1}(\phi_{B'})R_{y}^{-1}(\theta_{B'})=B'\hat{\boldsymbol{z}}.$
Then let $\theta_{w''}$ and $\phi_{w''}$ be the spherical angles
of $\boldsymbol{w}''$ in $\boldsymbol{r}''$-system. We do a final
rotation $\boldsymbol{r}'''  =  R_{z}(\phi_{w''})\boldsymbol{r}''$
to make $\boldsymbol{w}'''$ in the $x$-$z$ plane of the $\boldsymbol{r}'''$-system.

In summary, starting from an initial anisotropic model with arbitrary
$\boldsymbol{w}$ vector and $\boldsymbol{B}$-field, after we do
the coordinate transformation,
\begin{eqnarray}
\boldsymbol{r}''' & = & R_{z}(\phi_{w''})R_{y}(\theta_{B'})R_{z}(\phi_{B'})\left(\begin{array}{c}
x/\lambda_{x}\\
y/\lambda_{y}\\
z/\lambda_{z}
\end{array}\right),
\end{eqnarray}
we could map the original Hamiltonian to an isotropic model
\begin{eqnarray}
H_{0} & = & w_{x}^{\prime\prime\prime}k_{x}^{\prime\prime\prime}+w_{z}^{\prime\prime\prime}k_{z}^{\prime\prime\prime}+v_{0}
\boldsymbol{k}^{\prime\prime\prime}\boldsymbol{\cdot\sigma},
\end{eqnarray}
with a $\boldsymbol{B}'''$ filed along $z$-direction.
Drop the primes, we obtain the simple isotropic Hamiltonian, and then the LL results presented in the main text can be directly applied.

\section{LL Collapse in a Tight-binding Model }
Here we study the LL squeezing and collapse effect in a simple tight-binding (TB) model:
\begin{eqnarray}
\hat{H} & = & \sum_{i,j,k}c_{ijk}^{\dagger}V_{ijk}c_{ijk}+\sum_{ijk}\left(c_{i+1}^{\dagger}T_{x}c_{i}+c_{j+1}^{\dagger}T_{y}c_{j}+c_{k+1}^{\dagger}T_{z}c_{k}+\text{h.c.}\right)\label{eq:latH-1}
\end{eqnarray}
defined on a cubic lattice, where $c_{ijk}$ is the electron annihilation operator on the site ($i$, $j$, $k$), $V_{ijk}=w_z\cot k_{w}+\left(2+\cot k_{w}\right)\sigma_{z}$,
$T_{x(y)}=(i\sigma_{x(y)}-\sigma_{z})/2$, and $T_{z}=-(w_z+\sigma_{z})/(2\sin k_{w})$.
The corresponding Bloch Hamiltonian
in $k$-space is
\begin{eqnarray}
H(\boldsymbol{k}) & = & -w_z \frac{\cos k_{z}-\cos k_{w}}{\sin k_{w}}+\sin k_{x}\sigma_{x}+\sin k_{y}\sigma_{y}-\left(\cos k_{x}+\cos k_{y}-2+\frac{\cos k_{z}-\cos k_{w}}{\sin k_{w}}\right)\sigma_{z}.
\end{eqnarray}
There is a single pair of Weyl nodes located at $\boldsymbol{k}=(0, 0, \pm k_{w})$ with opposite chirality, and the low-energy Hamiltonian around the Weyl nodes reads,
\begin{eqnarray}
H(\boldsymbol{k}) & = & \pm w_{z}k_{z}+\sum_{i}v_{i}k_{i}\sigma_{i},
\end{eqnarray}
with $\boldsymbol{v}=(1, 1, \pm1)$. The tilt at the Weyl nodes is along $z$-direction.
When $|w_{z}|>1$ the Weyl node is type-II, while for $|w_{z}|<1$  the node is type-I.

\begin{figure}[h!]
\includegraphics[width=11cm]{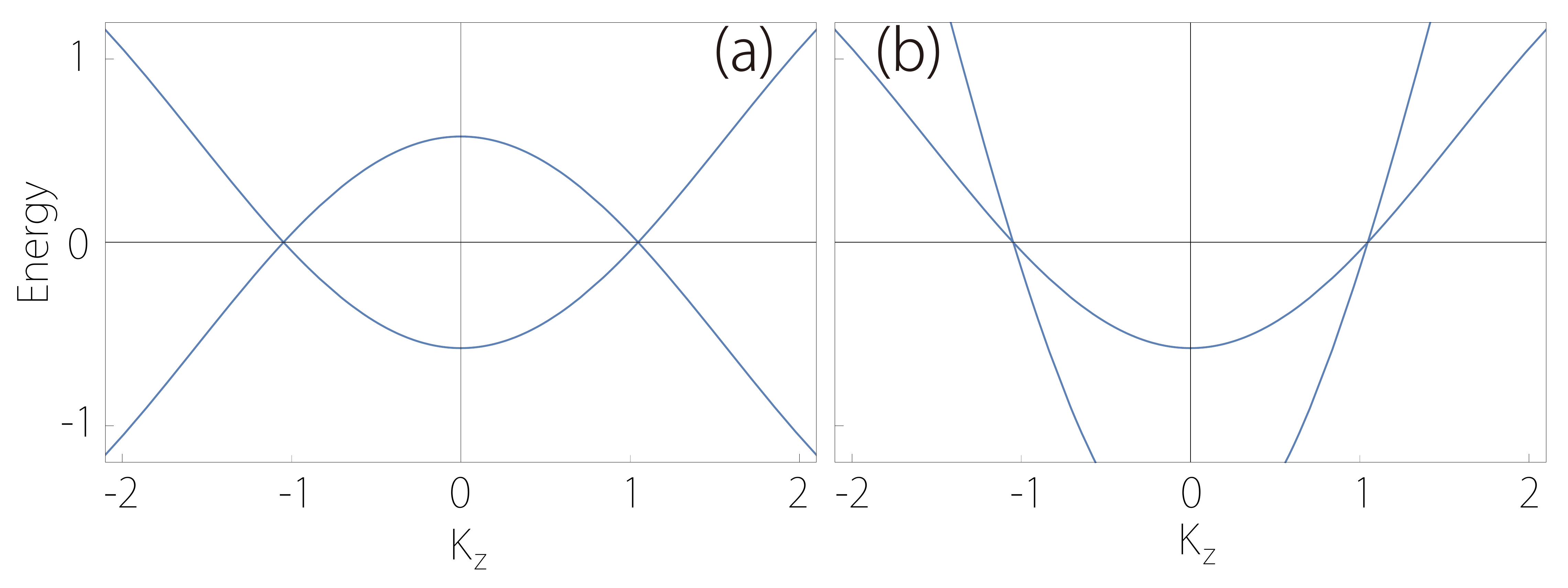}
\caption{Energy spectra for the tight-binding model, featuring two Weyl nodes along $k_z$-axis, with (a) $w_z=0$ and (b) $w_z=2$. Here, $k_w=\pi/3$, and we set $k_x=k_y=0$.}
\label{figS5}
\end{figure}

The magnetic field is simulated by the standard procedure of adding a Peierls phase in the hopping amplitudes. For a tight-binding model, it is difficult to continously vary the $B$-field direction with respect to the lattice, as what we did for the continuum model. Instead, we perform the following two sets of calculations.

We first consider the setup with $B$-field along $x$-direction, i.e. perpendicular to the tilt. Then we increase the magnitude of the tilt such that the $\beta$ parameter increases from 0 to 1. In Fig.~\ref{figS6}(a-c), we show the LL spectrum with $\beta=0,\ 0.8 \ \text{and}\  1$. It shows that with increasing $\beta$, the LL spacing is squeezed. At $\beta=1$, a huge number of LLs are collapse onto a single curve (the envelope curves near zero energy in the figure). In Fig.~\ref{figS6}(d), we plot the energies of the first few LLs with $k_x=0$, which clearly shows the the continuous squeezing and eventual collapse of the spectrum at $\beta=1$.

Next, we show the anisotropic field dependence. For fixed model parameters and $B$-field strength, we compare the LL spectrum for two different $B$-field orientations. In Fig.~\ref{figS7}(b), the field is along $z$-direction, i.e. along the tilt. Hence there is no LL collapse at the Weyl point and the LLs are well-separated. In contrast, for the field along $x$-direction, we have $\beta=1$ corresponding to the collapse point and indeed the spectrum is highly squeezed and the envelope curves around zero energy become highly degenerate.

\begin{figure}[H]
\centering
\includegraphics[width=13cm]{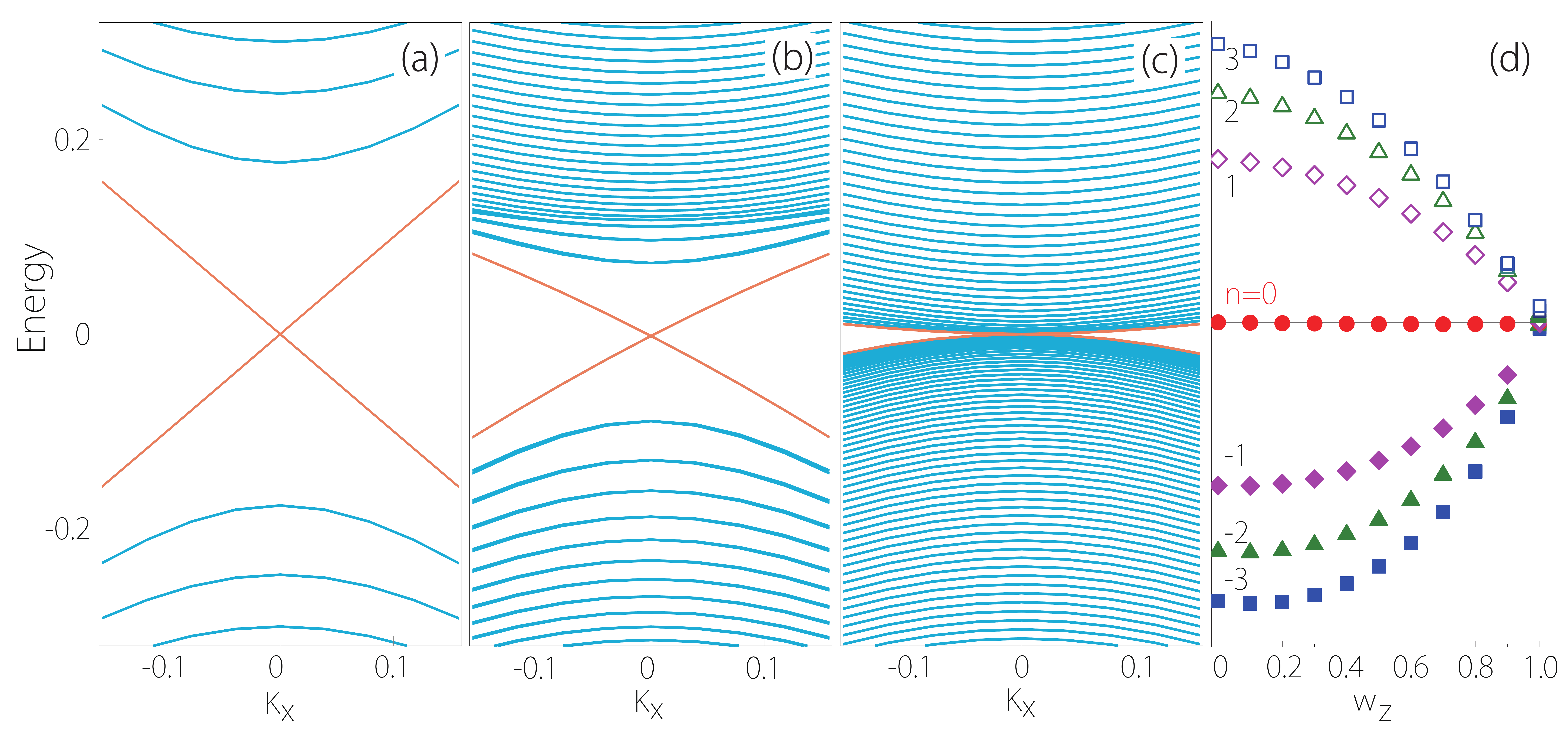}
\caption{LL spectra when $B$ field is along $x$-direction by varying the model parameter $w_z$, such that (a) $\beta=0$, (b) $\beta=0.8$, and (c) $\beta=1$. (d) Energies for the first few LLs (with $k_x=0$) plotted versus $\beta$. Here we set $k_w=\pi/3$.}
\label{figS6}
\end{figure}

\begin{figure}[H]
\centering
\includegraphics[width=9cm]{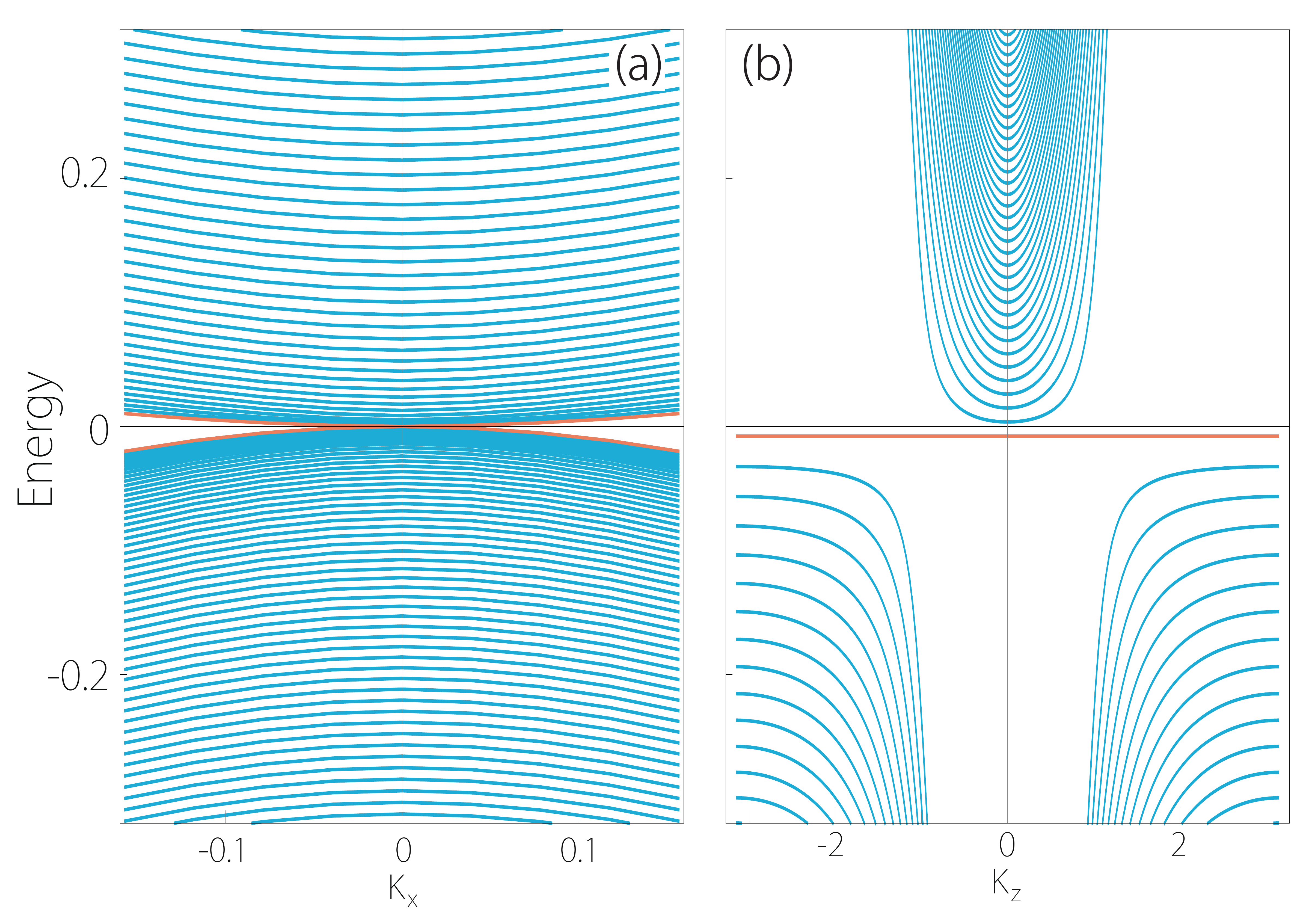}
\caption{The LL spectra for $B$-field (a) along $x$-direction, and (b) along $z$-direction. There is no LL collapse for case (b) ($\beta=0$), whereas collapse occurs for case (a) ($\beta=1$), showing the strongly anisotropic field dependence.}
\label{figS7}
\end{figure}
